\documentclass{aa} 

\usepackage{graphicx}
\usepackage{amssymb}
\newcommand{\etal}{et al.}
\newcommand{\vs}{vs.}
\newcommand{\eg}{e.g.}
\newcommand{\ie}{i.e.}
\newcommand{\cf}{cf.}
\newcommand{\muK}{~\mu{\rm K}}

\newcommand{\muKrs}{~\mu{\rm K}\sqrt{{\rm s}}}
\newcommand{\apj}{ApJ}
\newcommand{\PR}{Phys. Rev.}
\newcommand{\AsAs}{A\&A}
\newcommand{\AASS}{A\&AS}
\newcommand{\MNRAS}{MNRAS}

\begin{document}
\title{A Map-Making Algorithm for the Planck Surveyor}
 
\subtitle{}
\author{P. Natoli\inst{1} \and G. de~Gasperis\inst{1} \and C. Gheller\inst{2} \and N. Vittorio\inst{1}}

\institute{ Dipartimento di Fisica, Universit\`a di Roma ``Tor Vergata'', via della Ricerca Scientifica 1,
I-00133, Roma, Italy \and Cineca, Via Magnanelli 6/3, I-40033 Caselecchio di Reno (BO), Italy}

\offprints{Paolo.Natoli@roma2.infn.it}

\date{Received / Accepted}

%\markboth{P. Natoli \etal: A Map-Making Algorithm for the Planck Surveyor} {P. Natoli \etal: A 
%Map-Making Algorithm for the Planck Surveyor}

\abstract{
We present a parallel implementation of a map-making algorithm 
for CMB anisotropy experiments which is both fast and efficient. 
We show for the first time
a Maximum Likelihood, minimum variance map obtained by
processing the entire  data stream expected from the {\sc Planck} Surveyor, 
under the assumption of a symmetric beam profile. 
Here we restrict ourselves to the case of the 30 GHz channel of the
{\sc Planck} Low Frequency Instrument. The extension to {\sc Planck} higher 
frequency channels is straightforward.   
If the satellite pointing periodicity is good
enough to average data that belong to the same sky circle, 
then the code runs very efficiently on workstations.  
The serial version of our code
also runs on very competitive time-scales  
the map-making pipeline for current and forthcoming 
balloon borne experiments.   
}
\maketitle

\section{Introduction}\label{intro}
Making CMB maps out of balloon 
(\eg\ QMAP, de Oliveira-Costa \etal\ 1998;   
MAXIMA-1, Hanany \etal\  2000; BOOMERanG, de Bernardis \etal\  2000) 
or space borne (MAP\footnote{http://map.gsfc.nasa.gov/} and the {\sc Planck}
Surveyor\footnote{http://astro.estec.esa.nl/SA-general/Projects/Planck/}) 
experiments is
an important step of the standard data analysis pipeline. 
It allows for: a major lossless compression of
the raw data with a minimum number of assumptions; checks on
systematics; tests for and/or estimation of foreground contamination; etc.
Map-making is also important when simulating a CMB experiment. 
It allows to: look for
possible systematics; optimize the focal plane design and/or the experiment scanning strategy; etc. It is then highly desirable to develop tools able to attack the map-making problem also for
forthcoming, high-resolution CMB space experiments.

Map-making a la COBE (see \eg\ Lineweaver 1994) can be extended to the 
differential, high resolution  MAP experiment 
(Wright, Hinshaw and Bennett, 1996). Moreover Wright (1996) has
discussed how to perform map-making in the case of one-horned experiments significantly affected
by
$1/f$  noise, which introduces spurious correlations in the data time stream (see \eg\ Janssen \etal\ 
1996). To our knowledge this algorithm has not yet been implemented for the {\sc Planck} Surveyor
mission. To date, maps for {\sc Planck} have only been produced by means of {\it destriping} algorithms
(Delabrouille \etal\ 1998; Maino
\etal\ 1999). We remind that {\sc Planck} spins at 1~rpm, has a boresight angle of $85^\circ$ and
observes the same circle of the sky for 1 hour (every hour the spin axis is moved, say along the
ecliptic, by $2.5'$). The destriping algorithms implemented for {\sc Planck} assume that, after averaging
data taken along a given circle (\ie\ in one hour), the residual noise along the circle is well
approximated by white noise plus an offset. These offsets are determined by minimizing the set of all
possible differences between the antenna temperatures of the same sky pixel 
observed in two different
scan circles. This method rests on a number of assumptions, the stronger being that the knee
frequency of the $1/f$ noise is smaller than or at most comparable to the spin frequency.
Wright's method does not suffer this limitation because it assumes to
know the statistical properties of the noise,  directly derived from the data themselves (see \eg\
Ferreira and Jaffe, 2000; Prunet \etal\ 2000).

The purpose of this paper is to present the first implementation of 
Wright's method to the {\sc Planck}
Surveyor, for the moment assuming a symmetric beam profile. In fact, we want to
show, to our knowledge for the first time, the analysis of the entire time stream (14
months) of {\sc Planck} simulated data
to produce  Maximum Likelihood, minimum variance CMB maps (we stress
that maps obtained from destriping algorithms are not
necessarily minimum variance). 
The parallel, Message Passing Interface (hereafter MPI)
implementation of the algorithm has been  tested on a SGI Origin~2000
(hereafter O2K) and  runs on time
scales that might render Monte Carlo simulations feasible. 
This opens up the possibility of
evaluating the CMB angular power spectrum via Monte Carlo techniques 
(Wandelt, Hivon \& G\'orski 2000) rather than by maximizing a
Likelihood (see \eg\ Bond, Jaffe \& Knox 1998) or by directly evaluating the
map angular correlation function (Szapudi \etal\ 1999). 
We want to stress that our 
implementation does not assume any
exact periodicity of sky pointing within single circle observation 
(\ie\ no average on the circle is
performed, as instead 
required by destriping algorithms). 
If, on the contrary, it turns out that pointing
periodicity is a good assumption 
(\ie\ average on circles is performed) then the same code
runs very efficiently on medium sized workstations (\eg\ Pentium based PCs), again in quite short
time scales.

Although this paper might seem rather technical, 
we think that it can be of interest to a large
community of CMB data analysts, involved either in the {\sc Planck} collaboration or
 in the forthcoming, new generation balloon experiments.

The plan of this paper is as follows. In Sect.~2 and 3 we will briefly 
discuss the method and its
implementation. In Sect.~4 we will show tests and benchmarks of our implemented software. In
Sect.~5 we will apply our tools to {\sc Planck} and BOOMERanG simulated data.
Finally, in Sect.~6 we will briefly review our conclusions.

\section{Method}\label{method}

\subsection{Statement of the problem}\label{statement_of_prob} Let us
for completeness outline
here the map-making algorithm and its assumptions. The primary output of
a CMB experiment are the
Time Ordered Data (TOD), $\mathbf{d}$, which consist of $\mathcal{N}_d$
sky observations made
with a given scanning strategy and at a given sampling rate (three
points per FWHM, say). A map,
${\bf m}$, can be thought as a vector containing $\mathcal{N}_p$
temperature values, associated
with sky pixels of dimension 
$\sim$ FWHM/3. 
Following Wright (1996) we assume that the    TOD  depend linearly on
the map:
\begin{equation}
\label{measures} {\bf d} = {\bf P}{\bf m} + {\bf n},
\end{equation} 
where ${\bf n}$ is a vector of random noise and $\bf P$ is some known
matrix\footnote{Note that the elements of ${\bf m}$ need not to be sky
temperature values.
Other parameters may be fitted provided that they also depend linearly
on the TOD. Sometimes
these parameters are called ``virtual'' pixels.}. 
The rectangular,
$\mathcal{ N}_d \times
\mathcal{ N}_p$ matrix $\bf P$ is dubbed a {\em pointing matrix}.   That
is, applying $\bf  P$ on a
map ``unrolls'' the latter on a TOD according to a given scanning
strategy. Conversely, applying
${\bf P}^T$ on the TOD ``sums'' them into a map\footnote{The value of a pixel
of this map is the sum of all
the observations of that pixel made at different times according to a
given scanning strategy.}. The
structure of $\bf  P$ depends on what we assume for $\mathbf{m}$. If
$\mathbf{m}$ contains a
pixelized but unsmeared  image of the sky then $\mathbf{P}$ must account
for beam smearing. This
is the most general assumption and allows one to properly treat, for
instance, an asymmetric beam
profile. In this case, applying
$\mathbf{P}$ to $\mathbf{m}$ implies both convolving the sky pattern
with the 
detector beam
response and  unrolling
$\mathbf{m}$ into a ``signal-only'' time stream. If, on the other hand,
the beam is --at least
approximately-- symmetric then it is possible --and certainly more
convenient-- to consider ${\bf m}$
as the beam smeared pixelized sky. The structure of $\mathbf{P}$ for a
one-horned experiment would
then be very simple. Only one element per row would be different from
zero, 
the one connecting the
observation of $j$-th pixel to the $i$-th element of the TOD. Hereafter
we will restrict ourselves to
this case and we will address the implementation of an asymmetric beam
profile in a forthcoming
paper.

\subsection{Least Squares Approach to Map-making}\label{least_squares} 

Many methods have been
proposed to estimate ${\bf m}$ in Eq. (\ref{measures}) [for a review
see, \eg\ Tegmark (1997)].
Since the problem is linear in ${\bf m}$, the use of a Generalized
Least Squares (GLS) method appears well suited. This involves the 
minimization of the quantity
\[ 
\chi^2 = {\bf n}^T {\bf V} {\bf n} = ({\bf d}^T - {\bf m}^T {\bf P}^T)
{\bf V} ({\bf d} - {\bf P} {\bf
m})
\] for some nonsingular, symmetric matrix $\mathbf{V}$. Deriving with
respect to
${\bf m}$ yields an estimator, say ${\tilde{\bf m}}$, for the map:
\begin{equation}\label{mtilde} 
{\tilde {\bf m}} = ({\bf P}^T {\bf V} {\bf P})^{-1} {\bf P}^T {\bf V}
{\bf d}
\end{equation} 
The proof that this estimator is unbiased is straightforward. Just note
that:
\[ {\tilde {\bf m}} - {\bf m} = ({\bf P}^T {\bf V} {\bf P})^{-1} {\bf
P}^T {\bf V}{\bf n}.
\] So, provided that $\langle{\bf n}\rangle=0$, we have $\langle{\tilde
{\bf m}}\rangle = {\bf m}$
(the symbol $\langle\cdot\rangle$ indicates, as usual, an average 
over the ensemble). The map
covariance matrix is, then:
\begin{eqnarray*} 
{\bf \Sigma}^{-1} & \equiv &
\langle({\bf m}-{\tilde {\bf m}})({\bf m}^T-{\tilde {\bf m}}^T)\rangle =
\\ & = & ({\bf P}^T {\bf V} {\bf P})^{-1}{\bf P}^T {\bf V}\langle{\bf
n}{\bf n}^T\rangle  {\bf
V}{\bf P}({\bf P}^T {\bf V} {\bf P})^{-1}
\end{eqnarray*} 
In order to have a ``low noise'' estimator, one has to
 find $\mathbf{V}$ that minimizes the variance of ${\tilde {\bf m}}$.
This is attained if we take $\bf
V$ to be the noise inverse covariance matrix,
\ie\  ${\bf V}^{-1} = {\bf N} \equiv  \langle{\bf n}{\bf n}^T\rangle$.
Then
${\tilde\mathbf{m}}$ has the very nice property of being, amongst all
linear and unbiased estimators,
the one of   minimum variance 
\footnote{To see why this is the case, remember that any linear
estimator can be written in the form
$\hat {\bf m} = {\bf  Ad } = {\bf AP  m}  + {\bf A  n}$. So, the
condition $\bf AP = I$ must hold for
it to be unbiased. To prove that the variance of $\hat {\bf m}$ can not
be smaller than the variance
of ${\tilde {\bf m}}$, it helps (see, \eg\ Lupton 1993) to write
$\hat {\bf m} = {\tilde {\bf m}} +(\hat {\bf m} -{\tilde {\bf m}})$. It
follows that
\[ {\rm Var}(\hat {\bf m}) = {\rm Var}({\tilde {\bf m}})+ {\rm Var}(\hat
{\bf m} -{\tilde {\bf m}})+
2{\rm Cov}({\tilde {\bf m}},\hat {\bf m} -{\tilde {\bf m}}).
\] 
By direct substitution, one finds that
${\rm Cov}({\tilde {\bf m}},\hat {\bf m} -{\tilde {\bf m}})  =  ({\bf
TP-I})({\bf P}^T {\bf N}^{-1}
{\bf P})^{-1}  =  0$. Thus, the variance of $\hat {\bf m}$ is equal to
the variance of ${\tilde {\bf
m}}$ plus a nonnegative quantity, which is all we need to show. }. The
GLS
 solution to the map-making problem is, then:
\begin{equation}\label{map} {\tilde {\bf m}} = {\bf \Sigma}^{-1} \, {\bf
P}^T {\bf N}^{-1} {\bf d}
\end{equation} 
where
\begin{equation}
\label{cov} {\bf \Sigma} = {\bf P}^T {\bf N}^{-1} {\bf P}
\end{equation}

\subsection{ML Solutions} 

The
statistical properties of detector noise are, as usual, described by a
multivariate Gaussian distribution. This fact has two consequences. 
The first one is rather obvious: if
the noise distribution is Gaussian, ${\tilde {\bf m}}$ is indeed the
ML estimator. In
fact, in the Gaussian case, the Likelihood of the data time stream given
the (true) map is 
\begin{eqnarray}\label{likelihood} && \hskip-0.35truecm \mathcal{L}({\bf
d}|{\bf m}) =
(2\pi)^{-\mathcal{ N}_p/2}
\times \nonumber
\\ && \hskip-0.35truecm {\rm exp} \left\{ - {1 \over 2} [({\bf d}^T -
{\bf m}^T {\bf P}^T) {\bf
N}^{-1}
 ({\bf d} - {\bf
P} {\bf m}) + {\rm Tr}(\ln {\bf N})]
\right\} 
\end{eqnarray} Solving for the maximum would clearly give
Eq.~(\ref{map}) and Eq.~(\ref{cov})
again. The second remark has to do with the notion of a {\it lossless
map}, that is, a map which
contains all the relevant cosmological information contained in the TOD.
A good way to prove that a
method is lossless is to show that the TOD and the related map have the
same Fisher information
matrix.  The estimator defined in Eq.~(\ref{map})
leads to a lossless map (Tegmark 1997).

\section{Implementation} 

The method outlined in Sect.~\ref{method} has the advantage of being
simple,
linear and, hence, computationally appealing for {\sc Planck}. 
In this Section we discuss our implementation
of the map-making algorithm to the 30~GHz channel of the {\sc Planck} Low
Frequency Instrument (LFI),
with a nominal resolution $\sim 30'$ FWHM. This implies $\mathcal{N}_d
\sim 10^9$ (14 months of
observation,
\ie\  about two sky coverages) and $\mathcal{N}_p  \sim 10^6$. Note,
however, that the  
implementation we present here is  rather general since the algorithm
does not take specific
advantage of the details of {\sc Planck}'s scanning strategy and instrumental
performances. In fact, we
want to stress that our implementation of Wright's algorithm is, in
principle, well suited for any
one-horned, balloon- or space-borne CMB experiment.

\subsection{The Noise  Covariance Matrix}\label{Noise_CVM} 
As already mentioned in the Introduction, the GLS
method assumes to know the statistical properties of the noise. 
They are fully described by
$\mathbf{N}$,  the noise covariance matrix in the time domain. It is of
course highly desirable to
estimate $\mathbf{N}$  directly from the  data  since there is no
\emph{a priori} guarantee that the
noise statistical properties match  with what is measured during ground
testing. The measured
TOD is a combination of signal and noise. Thus, one has to  estimate the
signal, subtract  it from the
TOD and only then  estimate $\mathbf{N}$. 
Ferreira and Jaffe (2000) and Prunet \etal\ (2000)
discuss iterative methods to attack this issue. 

The problem of evaluating the noise properties directly from the TOD is
a bit far from the main
point we want to make here: the possibility, given $\bf N$, 
to analyze the entire {\sc Planck} simulated
data set to produce a ML, minimum variance map. In any
case, we will show in
Sect.~\ref{num_res} that for the {\sc Planck} case, 
to {\it zero}-th order, we can
estimate the signal by projecting the TOD onto the sky. This implies
applying
${\bf P}^T$ [summing the TOD into a map], $({\bf P}^T{\bf P})^{-1}$
[dividing the pixel values by
the number of hits] and ${\bf P}$ [unrolling a map into a time stream] to $\bf d$. 
The noise estimator, $\tilde {\bf n}$, can then be written as follows:
\begin{equation}\label{n_tilde} {\tilde \mathbf{n}} = \mathbf{d} - {\bf
P} ({\bf P}^T{\bf
P})^{-1}{\bf P}^T {\bf d} = [\mathbf{I}-{\bf P} ({\bf P}^T{\bf
P})^{-1}{\bf P}^T]\mathbf{n}
\end{equation} and does not include any contribution from the signal. 
 In a forthcoming paper (Natoli \etal\ ~2001) we will address in more
detail, 
and specifically for the {\sc Planck} Surveyor, the problem of estimating the 
noise statistical properties directly from flight data. 
Here, in Sect.~\ref{num_res}, we will only show that 
$\tilde \mathbf{n}$ allows a reasonably good estimate of the
``in-flight'' noise 
behavior.

\subsection{Stationary noise} 

It is customary to assume that the statistical properties of detector
noise do not change over the mission life time.  
The formal way to state
this property is to write the elements of the noise covariance 
matrix as follows:
\begin{equation} 
N_{ij} = {\xi}(|i-j|)
\end{equation} 
or, equivalently, to say that $\mathbf{N}$ is a
Toeplitz matrix. If $\xi$ vanishes for $|i-j| > \mathcal{N}_{\xi} \ll \mathcal{N}_d$, then
$\mathbf{N}$ is band diagonal and it can be well approximated
by a circulant matrix:
\begin{equation}\label{circ_def} N_{i+1,j+1} = N_{ij} \: \: \: \: \:
\rm{mod}\: \: \: {\mathcal N}_d.
\end{equation}
Obviously,  
circularity does not hold exactly for $\mathbf{N}$: there is no physical 
reason for the noise
correlation function to wrap around itself past the edges.
However, the number of elements to modify
in order to turn $\mathbf{N}$ into a circulant matrix is a mere
$\mathcal{N}_{ \xi}(\mathcal{N}_{\xi}+1) \ll \mathcal{N}_d^2$.
A circulant matrix is diagonal in Fourier space:
$\mathbf{N}=\mathbf{F}^\dagger{\bf \Xi}\mathbf{F}$, where $\mathbf{F}$
($\mathbf{F}^\dagger$) 
is the discrete Fourier transform (antitransform) operator (DFT), 
and $\mathbf {\Xi}$ is diagonal. 
It is trivial to show that the inverse of a circulant matrix is still 
circulant and, in particular, that
$\mathbf{N}^{-1}=\mathbf{F}^\dagger\mathbf{\Xi}^{-1}\mathbf{F}$.
Circularity ensures that we will not have edge problems: this is
exactly the boundary condition imposed by the DFT. 

\subsection{Non-stationary noise} 
The ML solution of the map-making problem [see
Eq.(\ref{map})   and Eq. (\ref{cov})] can be easily generalized to the
case of piecewise stationary
noise. In this case, the whole TOD can be split into a set 
$\{{\bf d}_{(\kappa)}\}$ of smaller pieces $\kappa = 1,\, \mathcal{N}_c$. 
Clearly, 
\begin{equation}
{\bf d}_{(\kappa)}= {\bf P}_{(\kappa)} {\bf m}+ {\bf n}_{(\kappa)},
\end{equation} 
where ${\bf P}_{(\kappa)}$ is a pointing matrix and ${\bf
n}_{(\kappa)}$ is a vector of random, stationary noise with covariance
matrix 
$ {\bf N}_{(\kappa)} $.  Neglecting
cross-correlations between different ${\bf n}_{(\kappa)}$'s (a
reasonable
assumption if $\mathcal{N}_{\xi} \ll \mathcal{N}_d$)
reduces the  Likelihood given in
Eq.(\ref{likelihood}) to the product of the Likelihood of the different
pieces, 
which of course do not need to share the same noise covariance matrix.
Thus, in this case, the
ML map is given by
\begin{equation}\label{sol_chunks} 
{\tilde\mathbf{m}} = \left[
\sum_{\kappa =1}^{\mathcal{N}_c}
\mathbf{P}^T_{(\kappa)}
\mathbf{N}^{-1}_{(\kappa)}\mathbf{P}_{(\kappa)}\right]^{-1}
 \sum_{\kappa =i}^{\mathcal{N}_c}
\mathbf{P}^T_{(\kappa)}\mathbf{N}^{-1}_{(\kappa)}\mathbf{d}_{(\kappa)}.
\end{equation} 
Since neglecting the correlation between different ${\bf
n}_{(\kappa)}$'s is 
\emph{de facto}
equivalent to treating different sections of the same TOD as independent 
experiment outcomes, we
can obviously think of  Eq.~(\ref{sol_chunks})  as a simple recipe to 
produce maps from
${\mathcal{N}_c}$  different horns whose sky coverage is at least
partially overlapped.

\subsection{The Noise Inverse Covariance Matrix}\label{Noise_ICVM} 
In what follows we do not need to explicitly use 
$\mathbf{N}^{-1}$, although we
have in principle all the relevant information to do so. In fact, the
first row of $\mathbf{N}^{-1}$,
usually called a noise filter in the literature, can be computed by
taking a DFT of ${\bf \Xi}^{-1}$, the
noise inverse spectral density. Then, by exploiting the circularity of
$\mathbf{N}^{-1}$ we could in principle reconstruct the whole matrix.
Needless to say, handling a
$\mathcal{N}_d \times \mathcal{N}_d$ matrix is, for the {\sc Planck} case,
unconceivable even for the 30~GHz case.

Fortunately, under the assumption of circularity the application of the
noise inverse covariance
matrix to the TOD is simply a convolution,
\begin{equation}\label{convolution}
\label{conv1} {\bf N}^{-1}{ \bf d} = \mathbf{F}^\dagger{\bf
\Xi}^{-1}\mathbf{F}{\bf d},
\end{equation} 
and we do not need to store a huge matrix.   

Asking how long a noise filter
should be is  a very important question. In fact, from the one hand, it
is
desirable to convolve the TOD with a filter properly sampled in 
$\mathcal{N}_d$ points (something already numerically feasible even for
a 
{\sc Planck}-like TOD). 
On the other hand,  Eq.~(\ref{convolution}) works for real data in the
limit 
$\mathcal{N}_\xi \ll \mathcal{N}_d$ (so that $\mathbf{N}^{-1}$ 
can be considered circulant).  
Also note that an obvious bound on $\mathcal{N}_\xi$ comes from the
instrument:
it is difficult to imagine that the {\sc Planck} detectors
will remain ``coherent'' for the
whole mission life-time. 
A further benefit in having $\mathcal{N}_\xi \ll \mathcal{N}_d$ is the
possibility of performing a piecewise convolution in
Eq.~(\ref{convolution}).
In any case, in Sect.~\ref{num_res}
we will show the sensitivity of our results to $\mathcal{N}_\xi$.

\subsection{The Conjugate Gradient Solver}\label{CG_solver}

Evaluating $\tilde \mathbf{m}$ requires solving Eq.~(\ref{map}), that for
convenience we rewrite as
\begin{equation}\label{linear_system} 
\mathbf{H}^{-1}{\bf P}^T  {\bf N}^{-1} ({\bf P}{\tilde\mathbf{m}}) = 
\mathbf{H}^{-1}{\bf P}^T  {\bf N}^{-1}  {\bf d}
\end{equation} 
where $ \mathbf{H}^{-1}$ is a preconditioning matrix (or preconditioner).
This linear system can be solved  by means of a Conjugate Gradient (CG) 
minimization technique (see \eg\ Axelsson \& Barker, 1984).
CG methods are known to converge efficiently [\ie\ in less than
the $\mathcal{O}(\mathcal{N}_p^3)$ {\bf operations} required by standard matrix
inversion]
if the eigenvalues of $\mathbf{\Sigma}$ 
spawn a few orders of magnitudes. If this is not the case, the method
should
be preconditioned. This implies finding a matrix
$\mathbf{H}$ such that its inverse approximates\footnote{To be quantitative, 
an optimal preconditioner would give 
$\mathbf{H}^{-1}\mathbf{\Sigma} = \mathbf{I} + \mathbf{R}$
with all the eigenvalues of $\mathbf{R}$ less than~1 
(Axelsson \& Barker, 1984). It is then clear that multiplying $\mathbf{\Sigma}$ 
by the preconditioner changes its spectrum
to give an equivalent but better conditioned system.} $\mathbf{\Sigma}^{-1}$.
The method then solves the equivalent system given in 
Eq.~(\ref{linear_system}).
Other than being a good approximation to the original matrix,
the preconditioner should also be fast to compute and invert.
These often are conflicting requirements: finding a good preconditioner
may be hard. In our case, we find more than satisfactory to 
precondition our system with the diagonal part of the symmetric
and positive definite matrix $\mathbf{\Sigma}$.
This choice is motivated
by the fact that $\mathbf{\Sigma}$ is diagonally dominant. 
Furthermore, the diagonal part of $\mathbf{\Sigma}$ is very close to
the number of hits per pixel $\mathbf{P}^{T}\mathbf{P}$ so in practice
we are
normalizing each row of the inverse covariance matrix to the
``redundancy'' of the corresponding 
pixel\footnote{One other problem should be mentioned: $\Sigma$
is remarkably ill conditioned. In fact,  any monopole term belongs 
to its kernel since applying  
${\bf N}^{-1}$ kills the corresponding TOD offset. Our preconditioner 
does not cure this problem which is instead fixed by
specifically  imposing that ${\tilde\mathbf{m}}$ has a
given 
(usually zero) average. This prescription, critical if one attempts
a matrix inversion, is not needed in our case: we can, of course,
always subtract out the monopole.}.

A CG algorithm does not need to explicitly
invert $\mathbf{\Sigma}$.  
We want to stress that
earlier methods implemented to solve the map-making problem for 
balloon-borne experiments
(Borrill, 1999) require the inversion of the same matrix, surely an
unpleasant task when considering
large maps. Such implementations usually rely on the fact that $\mathbf{\Sigma}$ is 
(analytically  speaking) positive definite and, as such, a 
candidate to be Cholesky decomposed in 
$\mathcal{O}(\mathcal{N}_p^3)$ operations. This fact makes
the procedure prohibitively
expensive --even for present day supercomputers-- when $\mathcal{N}_p$
becomes large ($\gtrsim 10^5$). Also, $\mathbf{\Sigma}$ 
may not be exactly positive definite due to small inaccuracies when
estimating the noise correlation properties, making Cholesky
decomposition critical. Note that using an iterative solver could
potentially reduce the above operation count by a factor 
$\mathcal{N}_p/\mathcal{N}_{iter}$. However, storage requirement for $\mathbf{\Sigma}$
would still be $\mathcal{O}(\mathcal{N}_p^2)$, prohibitive for {\sc Planck}

To conclude this Section, let us stress that the map we obtain, 
$\tilde \mathbf{m}$, has the correct noise covariance matrix,
$\mathbf{\Sigma}$, 
even if we never evaluate nor store it directly.
\subsection{Parallelization}
The scheme outlined above naturally lends itself to a multi-processor
environment. In fact, it is straightforward to divide the entire TOD
in a set of smaller, partially overlapped, time streams and force
each of the Processor Elements (PE) to perform map-making with its
own time stream. This is accomplished by explicitly assigning work
loads
to each PE by means of MPI calls embedded in the Fortran~95 code.

As stated above, the working assumption that validates
Eq.~(\ref{convolution})
is that $\mathcal{N}_\xi \ll \mathcal{N}_d$. Since the number of PE,
$\mathcal{N}_{PE}$, is
at most $\mathcal{O}(100)$ the previous condition is even fulfilled for
the 
single PE. Thus, each PE performs the convolution of
Eq.~(\ref{convolution})
by executing its own serial FFT as discussed in
Sect.~\ref{Noise_ICVM}. 
This is more efficient than spreading 
over the PE's an intrinsically parallel FFT, the advantage being that 
we limit inter-processor communications. However, as a drawback, each PE
has to handle a  partial map of size $\mathcal{N}_p$ and
cross talk
among PE's is necessary to merge $\mathcal{N}_{PE}$ of these maps at the
end
of each CG iteration. Note that 
increasing the number of processors 
$\mathcal {N}_{PE}$ while keeping $\mathcal {N}_d$ constant may 
result in the operation count being dominated by this merging (see
however
Sect.~\ref{scaling_NPE} below). 
 
\section{Performances} 
In this Section we benchmark an implementation
of the map-making algorithm described above. 
The benchmarks are produced
within the framework of the {\sc Planck}'s simulation pipeline. However, since
the algorithm does not take
specific advantage of {\sc Planck}'s scanning strategy and instrumental
performance, they are indeed
very general. To show that this is the case we will apply our
code to a simulation of the BOOMERanG experiment 
(see Sect.~\ref{Boom_map} below).

\subsection{Scaling with ${\mathcal N}_\xi$, ${\mathcal N}_d$ and 
${\mathcal N}_p$}\label{sec:scalinglaws}

Part of the {\it l.h.s.}\ of Eq.(\ref{linear_system}) can be thought as 
the application of the ${\bf P}^T{\bf
N}^{-1}$ operator (the same acting on $\mathbf{d}$ on the {\it r.h.s.})
to 
${\bf P}{\tilde{\bf m}}^{(k)}$, the data stream obtained by unrolling
the 
temptative solution produced by the CG at
the $k$-th step. While the application of ${\bf P}^T{\bf  N}^{-1}$ to
$\mathbf{d}$ is done just once, the application of the same operator to
${\bf P}{\tilde{\bf m}}^{(k)}$ must be done for every iteration of the
CG algorithm. 
Therefore, speed of execution becomes critical. 

\begin{figure}
\resizebox{\hsize}{!} {\includegraphics{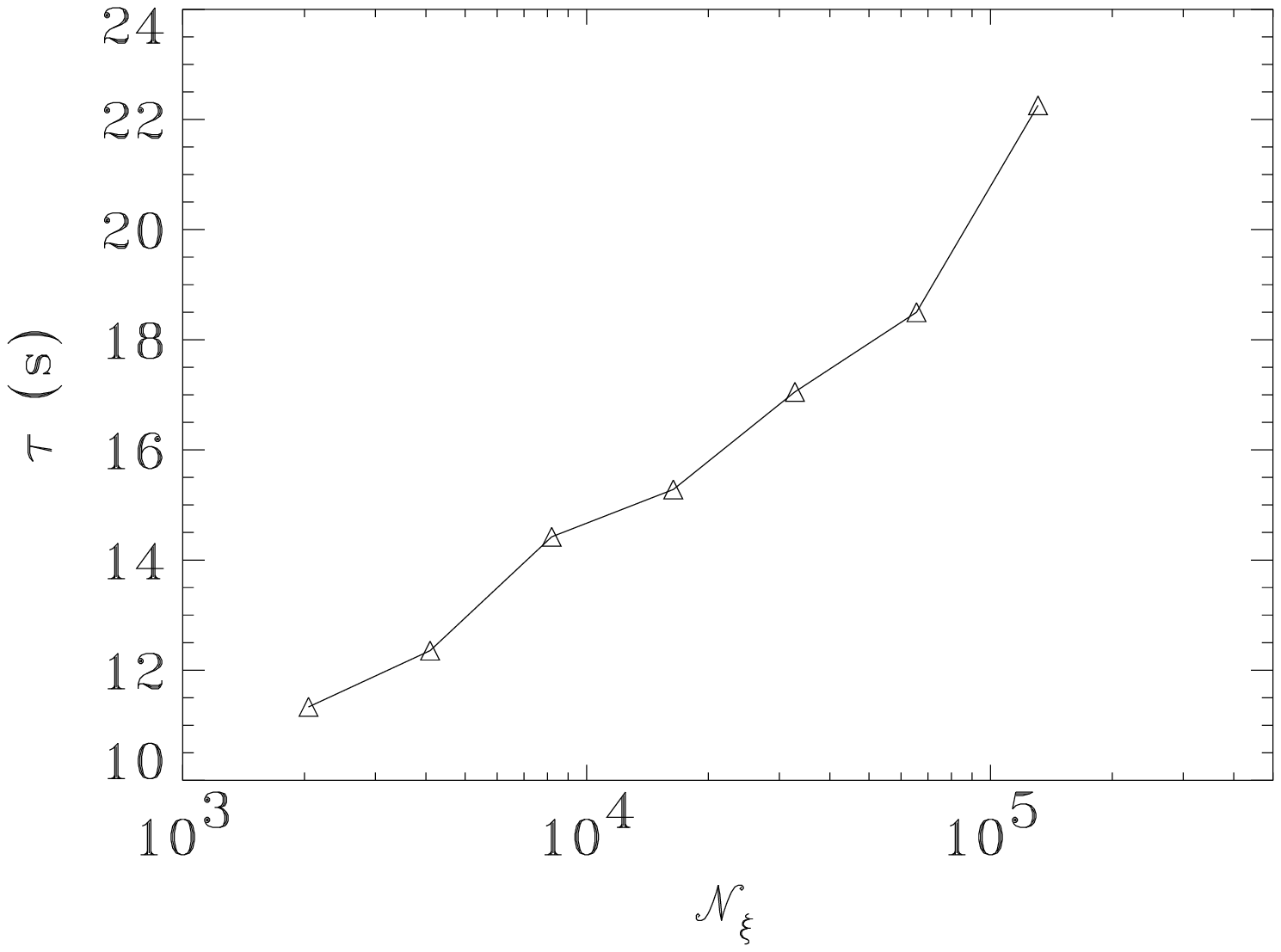}}
\resizebox{\hsize}{!} {\includegraphics{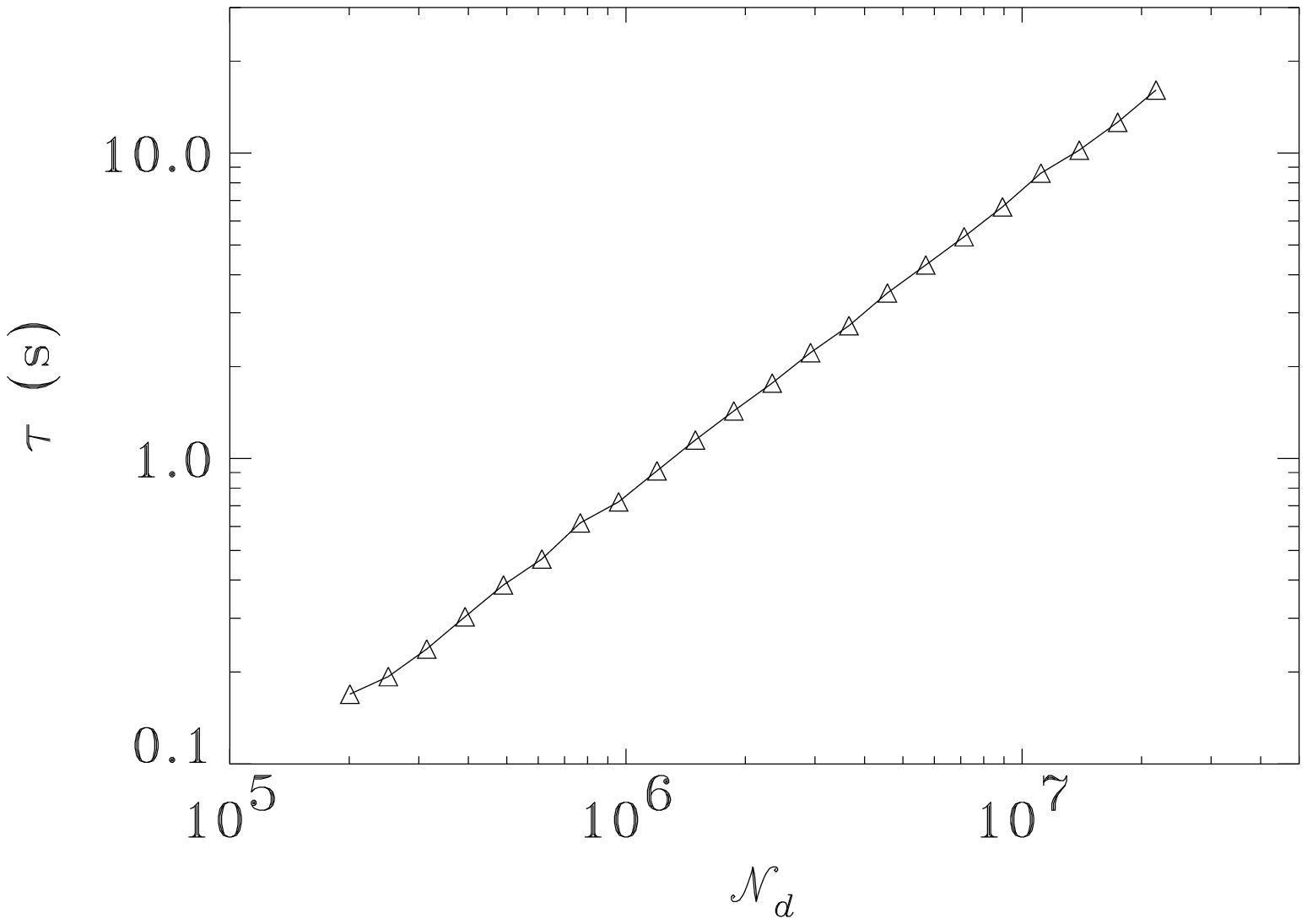}}
\resizebox{\hsize}{!}{\includegraphics{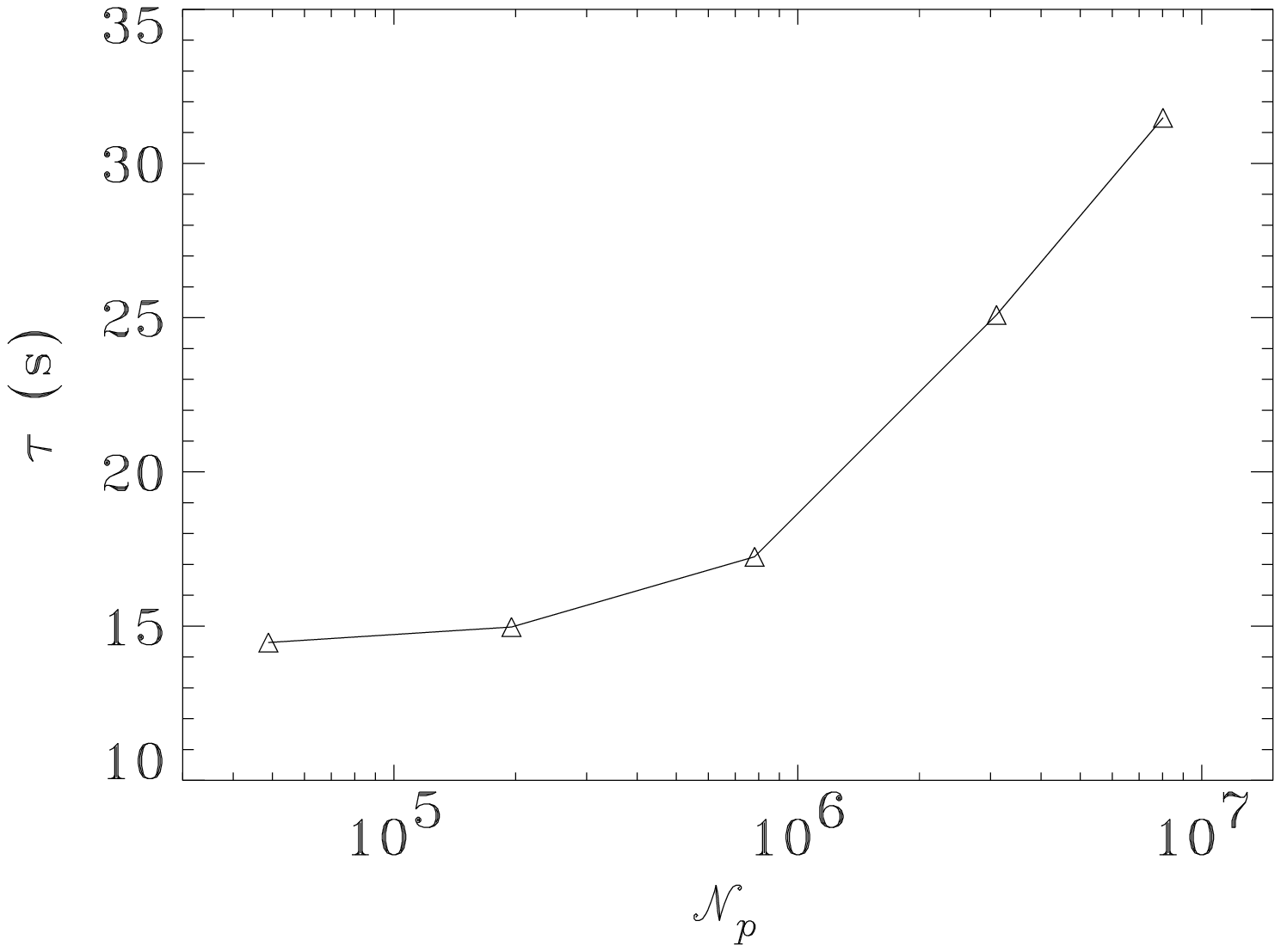}}
\caption{The wall-clock time $\tau(s)$ on a single 500~MHz Pentium III 
CPU  per CG iteration plotted \vs\ the noise filter length,  
$\mathcal{N}_\xi$, \vs\ the TOD length, $\mathcal{N}_d$  and \vs\ the map
pixel number, $\mathcal{N}_p$.}
\label{scaling_all}
\end{figure}

If we use a radix-2 Fast Fourier Transform (FFT) to perform a piecewise 
convolution of the TOD, the operation count is expected 
$\propto \mathcal{N}_d \log_2 \mathcal{N}_\xi$.
The lion share of CPU time is taken by
the (real) DFT transform. The use of an efficient FFT library greatly
speeds up
the calculation. 
We used the publicly available FFTW library (Frigo \& Johnson 1998). 
This code has
the nice property of tailoring itself to the architecture over which is 
executed, therefore greatly
enhancing cross-platform efficiency. In Fig.~\ref{scaling_all} (upper
panel) 
we plot CPU time per CG iteration, $\tau$, versus $\mathcal{N}_{\xi}$.
The behavior shown in the figure is due to the non constant
efficiency attained by the FFTW library when performing transforms of 
different lengths. 
FFT routines are usually more efficient when processing power-of-two
transforms and FFTW is no exception. 
Thus, we fix $\mathcal{N}_{\xi}$ to a power of two.
 
The application of ${\bf P}$
(${\bf P}^T$) to a map (TOD) is expected to scale linearly with
$\mathcal{N}_d$. In Fig.~\ref{scaling_all} (middle panel) we give $\tau$
as a
function of $\mathcal{N}_d$  for a given map size 
[$\mathcal{N}_p = 786\,432$, as expected by choosing 
$\mathcal{N}_\mathrm{side}=256$ in the HEALPix 
pixelization (G\'orski \etal\ 1999)] 
and the {\sc Planck} baseline scanning  strategy. 
The scaling with $\mathcal{N}_d$ is indeed linear:
$\tau= [(\mathcal{N}_d/1,400) + 50]$ms.

Since we never build up the ${\mathcal N}_p \times {\mathcal N}_p$ 
inverse covariance matrix $\mathbf
{\Sigma}$, 
but rather perform the piecewise multiplication
${\bf P}^T{\bf N}^{-1} ({\bf P}{\bf m})$
each time we need to do so, this operation is not 
expected to scale with the number of pixels.
However, multiplying by the preconditioner, $\mathbf{H}^{-1}$, does.
When ${\mathcal N}_p \ll {\mathcal N}_d$ (a condition usually fulfilled
for a CMB experiment) the preconditioner
should produce a negligible effect on the operation count. 
Conversely, we expect an increase in the operation count
as, keeping ${\mathcal N}_d$ fixed, ${\mathcal N}_p$ increases.   
This is confirmed in Fig.~\ref{scaling_all} (lower panel).
Note that, for the last two points in the plot, $\mathcal{N}_p \sim \mathcal {N}_d$.
Another factor that should in principle contribute to
the scaling of $\tau$ with ${\mathcal N}_p$ is the performance
overhead resulting from handling the arrays containing the maps. 
Note, however, that $\tau$ increases by only a factor $\sim 2$ when
${\mathcal N}_p$ is boosted by a factor $> 200$.

All the scalings given in Fig.\ref{scaling_all} have been obtained
from a single processor job using a 500~MHz Pentium III CPU.
\subsection{Scaling with $\mathcal{N}_{PE}$}\label{scaling_NPE}
In order to test our parallel software we use the following 
simulated data sets (relative to the LFI 30~GHz channel),
for which $\mathcal{N}_d$ changes by a couple of orders of magnitudes: 
$\mathcal{P}$60, the time stream obtained by averaging on circles 
the full {\sc Planck} TOD ($\mathcal{N}_d = 20,196,000$); $\mathcal{P}$3, 
a fictitious time stream obtained assuming
that each circle is observed only 20 times 
($\mathcal{N}_d = 403,920,000$);
  $\mathcal{P}$,  the full {\sc Planck} TOD ($\mathcal{N}_d =1,211,760,000$). 
In the ideal case, the algorithm speed $\tau^{-1}$ 
should scale linearly with $\mathcal{N}_{PE}$ and should be
inversely proportional to  $\mathcal N_d$. So, we can define an
efficiency index
%%%
\begin{equation}\label{epsilon}
  \mathcal{E} = 
  \frac{1}{\tau} \,\frac{\mathcal{N}_d}{\mathcal{N}_{PE}}  
 % \tau\,\frac{\mathcal{N}_{PE}}{\mathcal{N}_d}
\end{equation}
%%%
that we measure on an O2K.  For $\mathcal{P}$60, 
increasing $\mathcal {N}_{PE}$ from $1$ to $8$
results in  a loss of efficiency of $\sim 50\%$ due to inter-processor communications.
If we keep $\mathcal {N}_{PE}=8$
but change $\mathcal{P}$60 to $\mathcal{P}$3 and $\mathcal{P}$, 
$ \mathcal{E}$ increases. We find that
$\mathcal{E}$ is basically the same when comparing
$\mathcal{P}$60 and 
$\mathcal {N}_{PE}=1$ with $\mathcal{P}$ and $\mathcal {N}_{PE}=8$. Thus,
for these cases, our code shows the ideal scaling given in Eq.~\ref{epsilon}.

\begin{figure}
\resizebox{\hsize}{!}{\includegraphics{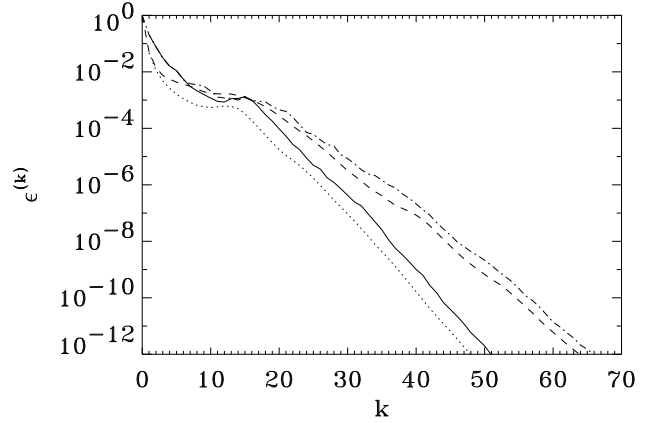}}
\caption{Solver precision as a function of the CG iteration step.
Solid and dotted curves refer to the
parallel code running on the full $\mathcal {P}$ TOD 
(no average on circles) for
``signal-only'' and ``noise-only'' TOD, respectively. Dot-dashed and
dashed curves refer to the serial code running on $\mathcal {P}60$, again
for ``noise-only'' and ``signal-only.''}
\label{solverprec}
\end{figure}
\subsection{Convergence criterion and precision}\label{conv_prec}
A natural criterion to stop the CG solver is to require 
that the fractional difference 
\begin{equation} \label{eq:epsilon_precision}
\epsilon^{(k)} = \frac{||{\bf P}^T {\bf N}^{-1}{\bf P}
  {\tilde\mathbf{m}^{(k)}} - {\bf P}^T {\bf N}^{-1}\mathbf{d}||} {||
  {\bf P}^T{\bf N}^{-1}\mathbf{d}||}
\end{equation}
obtained at the $k$-th iteration step is less than a supplied
tolerance. The norm $||\cdot||$ in Eq.~(\ref{eq:epsilon_precision})
represents the usual Euclidean norm\footnote{An alternative convergence
criterion is the more stringent $L_\infty$ norm, \ie\ the absolute value 
of the largest element of a vector. We have tested
that, for our applications, the CG solver returns almost identical
results
using either the Euclidean or $L_\infty$ norm. However, in the
latter case, convergence is significantly slower.}.

To test the CG convergence efficiency we plot 
in Fig.~\ref{solverprec} $\epsilon^{(k)}$ for two datasets of different
length: $\mathcal{N}_d = 20,196,000$ (corresponding to $\mathcal{P}$60)
and $\mathcal{N}_d =1,211,760,000$ (corresponding to $\mathcal{P}$).
Both
datasets have been benchmarked for the cases of ``signal-only'' and
``noise-only'' TOD. In the latter case, the noise properties
are kept independent of the TOD length to facilitate comparisons.
Note that, after a
few steps, the precision $\epsilon^{(k)}$ decreases exponentially and 
drops to the $10^{-6}$ level in
a few tens of iterations. Although always exponential, the
rate of convergence of the CG solver depends on the TOD length.
This is a consequence of having kept fixed $\mathcal{N}_\xi$ and $\mathcal{N}_p$
while increasing $\mathcal{N}_d$. In fact, the number of elements of
$\mathbf{\Sigma}$ connected
by the noise filter is lower for a longer TOD, a fact that
makes this matrix more efficiently preconditioned by its diagonal
part.
 
To quote a single number, the algorithm converges, to $10^{-6}$
precision,
in about 90 minutes ($\mathcal {N}_{PE} = 8$) to produce a map out
of the entire (\emph{not} averaged over circles) 14 months time stream,
expected from the {\sc Planck} Surveyor @ 30~GHz.

Of course, solver precision and accuracy (that is, ``distance'' from
the true solution) are not the same thing. In other words, being
confident that the algorithm converges quickly is only half the story.
We also want it to converge to the exact solution.
This is a rather important issue because we would not like the algorithm
to kill modes in the map. Discussion of this point is deferred until
the next Section.
\section{Numerical results}\label{num_res}
\subsection{``Signal-only''}\label{sigmaps}
In the limiting case of a noiseless TOD (\ie\ $\mathbf{d}=
\mathbf{P}\mathbf{m}$) Eq.~(\ref{map}) and Eq.~(\ref{cov}) yield
${\tilde \mathbf{m}} =
\mathbf{m}$. Thus, any good map-making code should return the ``true'',
pixelized sky when fed with
a ``signal-only'' TOD. We tested our code with  a pure CMB
anisotropy pattern \footnote{We use the standard Cold Dark Model with
$\Omega_{CDM}=0.95$, $\Omega_B=0.05$  and
$h=0.5$. No dipole was included.} 
and the same CMB pattern contaminated by the
foregrounds most relevant at 30~GHz~\footnote{Specifically, synchrotron
and free free emission  were included as well as simulated 
emission from SZ clusters and millimetric bright sources.}. The following
scheme  was employed to prepare the
TOD. First, a ``signal map'' is built, 
pixelized at $\sim$ FWHM/3 using the HEALPix 
pixelization ($\mathcal{N}_\mathrm{side}=256$) and FWHM-smeared. 
The latter is ``read'' by simulating a 30~GHz {\sc Planck}/LFI time stream. 
This TOD is the input to the map-making code. 

\begin{figure}
%\resizebox{\hsize}{!}{\includegraphics[angle=-270]{h2645f05.eps}}
\resizebox{\hsize}{!}{\includegraphics[angle=0]{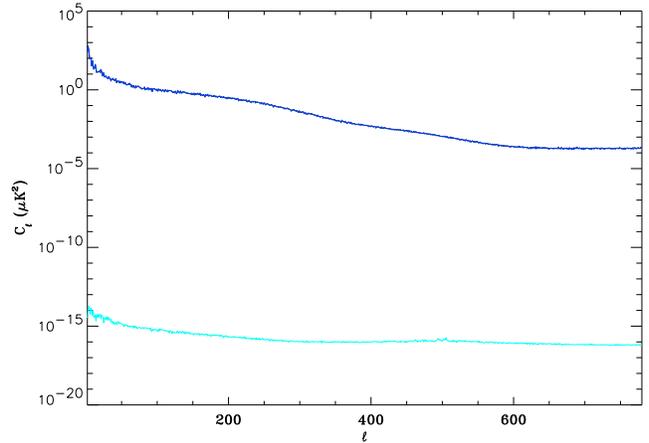}}
\caption{Upper panel: map-making output on ``signal-only'' TOD (CMB); 
lower panel:
angular power spectrum of the input map (dark blue line),
together with the spectrum of the map given in upper panel. The two are
virtually indistinguishable. The light blue line below shows the 
spectrum of the difference map.}
\label{map:sigonly}
\end{figure}

In order for the code to run a noise filter, $\mathbf{N}^{-1}$, 
is needed. First note that for this noiseless case the 
filter should, in principle, be irrelevant: as stated above, 
the GLS solution trivially gives back the ``true'' map 
independently of $\mathbf{N}^{-1}$. 
We must keep in mind, however, that we only consider $\mathcal{N}_\xi$
elements of the filter. In Fourier space this translates into estimating
the noise inverse spectral density precisely at $\mathcal{N}_\xi$ frequencies, 
the lowest, $f_\mathrm{min}$ say, being $\mathcal{N}_\xi$ times smaller 
than the sampling frequency.
%
% {\bf If the sky is observed for $T = n$ years (n being an integer) then
%the signal lives at the integer multiples of the spin frequency $f_\mathrm{spin}$
% (and at two frequencies disting $\pm 1/T$ from each multiple
% of $f_\mathrm{spin}$). }
%
One obvious limitation on $\mathcal{N}_\xi$ is that it must be 
fixed accordingly,
so that the Fourier representation of $\mathbf{N}^{-1}$ well comprises
the spin frequency. 
% {\bf Since the life time of the {\sc Planck} mission is not
%expected to be an integer multipole of one year a tiny amount of signal
%leaks into the lowest frequencies.}  
The amplitude of modes with $f <f_\mathrm{min}$, if any, is not recovered
unless the corresponding frequencies are considered in the analysis. 
However, there are
two reasons why one should not worry about this effect: first, and more
importantly, it is small and becomes utterly negligible for $\ell 
\gtrsim$ a few. Furthermore, as
previously stated, this contribution is likely to get lost in any
case since it falls in a region of the time Fourier spectrum 
strongly dominated by the noise 
and/or, more realistically, suppressed by hardware-induced decoherence 
(\eg\ instrumental high passing).
We show (for the typical case 
$\mathcal{N}_\xi = 4096$) the angular power spectra of the CMB only 
input and output maps together with their difference (lower panel of 
Fig.~\ref{map:sigonly}). It is clear that our map-making 
code recovers, as expected, the input power
spectrum to an impressively high
precision over the whole range of explored multipoles (although with
a somewhat coarser accuarcy at low $\ell$'s). 
The same happens when considering the case of CMB plus foregrounds 
(see Fig.~\ref{map:fullsig}, upper panel): the level of the residuals between the
input and output map is remarkably low, underlying the
high accuracy of our code  in recovering the input signal even in the
presence of sharp features (see Fig.~\ref{map:fullsig}, lower panel).

As shown above, the code also recovers to very high precision
even the broadest features in the map, \ie\ the ones corresponding
to low $\ell$'s. We want to stress that the CG solver converges to
the corresponding modes in a very reasonable number of iterations 
($\lesssim 50$, see Fig.~\ref{solverprec}). Other map-making codes,
implementing different solving algorithms, seem not to share
this property (Prunet \etal\ 2000) requiring more sophisticated methods
to speed up the convergence of low multipoles and/or a change
of variable to solve for the featureless noise-only map (see Dor\'e \etal\ 2001). 

\begin{figure}
%\resizebox{\hsize}{!}{\includegraphics[angle=-270]{h2645f07.eps}}
%\resizebox{\hsize}{!}{\includegraphics[angle=-270]{h2645f08.eps}}
%\resizebox{\hsize}{!}{\includegraphics[angle=0]{diff_spe_fullsig.ps}}
\caption{Upper panel: map-making output on ``signal-only'' TOD (CMB and main 
foregrounds). Lower panel: Difference between the above and input maps.
}
\label{map:fullsig}
\end{figure}

\begin{figure}
\resizebox{\hsize}{!}{\includegraphics[angle=0]{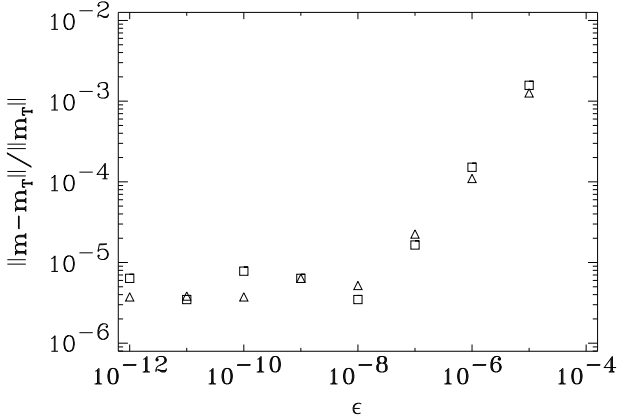}}
\caption{Maximum deviation between input and output maps in the 
``signal-only'' case, as a function of the solver precision.
Here $\mathcal{N}_\xi = 2^{11}$ (triangles) and $\mathcal{N}_\xi = 2^{17}$
(squares).}
\label{sigmaxprec}
\end{figure}

A final comment before concluding this Section. As mentioned at the end 
of Sect.~\ref{conv_prec}, precision and accuracy are not the same thing. 
We want to be sure that the CG algorithm converges to the
``true'' solution. So, from a given ``signal-only'' 
TOD we generate different maps by changing the
solver precision. 
For each of these maps we evaluate the
maximum difference in pixel value between them and the input map. The results are shown in Fig.~\ref{sigmaxprec} where we plot this 
maximum difference, normalized to
the $L_\infty$ norm of the input map, as a function 
of the required solver precision $\epsilon$. 
There are two regimes separated by $\epsilon \sim 10^{-8}$:  
the map accuracy decreases when  $\epsilon$ gets larger and 
saturates when $\epsilon$ gets smaller. 
So, $\epsilon\sim 10^{-8}$ is just about the right 
number to have a good compromise between speed of 
convergence and accuracy. In any case, any residual map error
smaller than $\sim 1\muK$ is acceptable in most applications, given the
{\sc Planck} sensitivity goal. 

%Dependence on $\mathcal{N}_\xi$

\subsection{"Noise-only"} 

In the opposite limiting case of a ``noise-only'' 
TOD (\ie\ $\mathbf{m} \equiv 0$), 
the map-making solution [\cf\ Eq.(\ref{map}) and
Eq.(\ref{cov})] gives
${\tilde \mathbf{m}} = \mathbf{(P^TN^{-1}P)^{-1}P^TN^{-1}n}$. 
The efficiency of a map-making 
code can be tested, in this case, by examining the
quality of the map obtained, 
a natural figure of merit being the variance of the reconstructed
map or -better- its angular power spectrum. 

We generate a noise time stream assuming the following form 
for the noise spectral density:
\begin{equation}
\label{pdf} P(f)=A[1+(|f|/f_k)^\alpha], 
\end{equation}
where $f_k$ is the  knee frequency. We choose  
$f_k=0.1$~Hz (the {\sc Planck} goal), $\alpha=-1$ and an amplitude A
corresponding to the expected white noise level of the $30$~GHz 
receivers. The minimum and maximum
frequencies are set by the inverse of the mission life-time 
and the sampling frequency ($33$~Hz).  
We then generate a ``noise-only'' data stream that
is the input to the map-making code.  
 
\begin{figure}
%\resizebox{\hsize}{!}{\includegraphics[angle=-270]{h2645f10.eps}}
%\resizebox{\hsize}{!}{\includegraphics[angle=-270]{h2645f11.eps}}
\resizebox{\hsize}{!}{\includegraphics[angle=0]{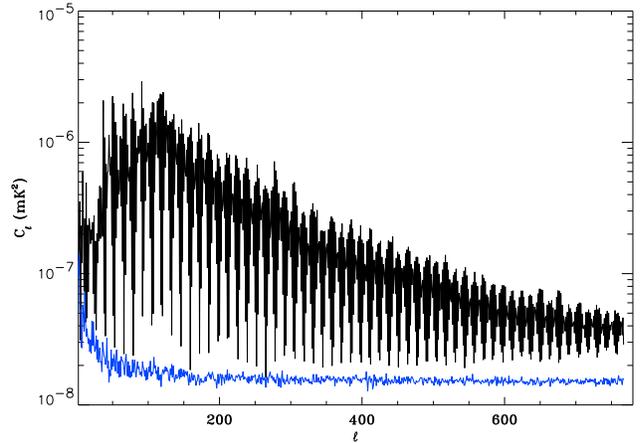}}
\caption{Upper panel: coadded map made out of a ``noise-only'' TOD. 
Note the stripes.
Middle panel: map-making solution on ``noise-only'' TOD.
Lower panel: angular power spectrum of the maps given above (black and
blue curves, respectively).
}
\label{map:coaddednoise}
\end{figure}

Fig.~\ref{map:coaddednoise} shows
the  map obtained by just averaging different 
observations of the same pixel (upper panel), 
the ML solution (middle panel) and the angular 
power spectra of the two (lower panel). 
It is clear from the middle panel of this figure that 
the map-making algorithm strongly suppresses the stripes due to
$1/f$ noise, very visible in the upper panel.
This is confirmed by the angular power spectrum of the ML
map (see lower panel of Fig.~\ref{map:coaddednoise}) which is basically
flat (as expected in the case of white and isotropic noise) for $\ell \gtrsim 100$.
The increasing power at $\ell \lesssim 100$ is due to two effects:
residuals
of spurious correlations in the noise and nonuniform sky 
coverage due to the {\sc Planck} scanning strategy. 
In spite of that, the overall level of 
instrumental noise has been considerably lowered by the map-making
algorithm.

\begin{figure}
%\resizebox{\hsize}{!}{\includegraphics[angle=0]{cl_teo-til.ps}}
\resizebox{\hsize}{!}{\includegraphics[angle=0]{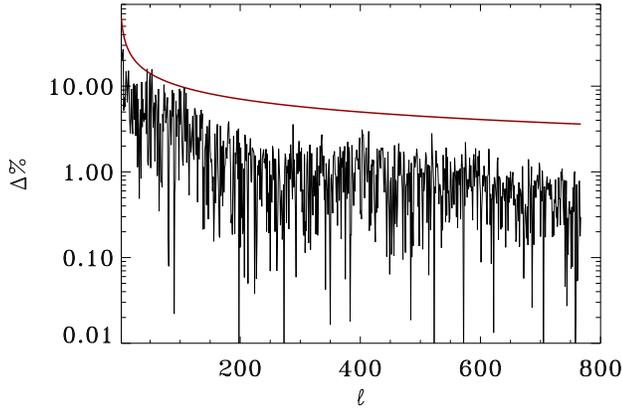}}
\caption{
Percentual difference, as a function of $\ell$, of the power spectra 
of the maps obtained using the true noise properties and
the noise estimator ${\tilde \mathbf{n}}$ discussed in the text.
}
\label{spe:teo-til}
\end{figure}

The noise filter or, equivalently, the inverse of the 
noise spectral density, has been evaluated as
explained in Sect.~\ref{Noise_CVM}. 
In order to assess how good $\tilde{\mathbf n}$ is as a noise estimator, 
we use the same ``noise-only'' TOD to produce two maps: 
the first, by using the ``true'' theoretical noise spectral
density [\cf\ Eq.(\ref{pdf})]; the second, 
by using the noise spectral density
estimated with $\tilde{\mathbf n}$. 
In both cases $\mathcal{N}_\xi = 4096$ 
as for the map in Fig.~\ref{map:coaddednoise}. 
In Fig.~\ref{spe:teo-til} 
we show, as a function of $\ell$, the
percentage difference between the spectra
derived from the two maps. 
So, although not optimized to the
specific case of the {\sc Planck} Surveyor, the noise 
estimator $\tilde{\mathbf n}$ reproduces to
quite a good extent the in-flight noise properties. 
In any case,
as also shown in Fig.~\ref{spe:teo-til}, the uncertainties
introduced in the estimate of the noise angular power spectrum are 
less than those due to cosmic variance.

\begin{figure}
\resizebox{\hsize}{!}{\includegraphics{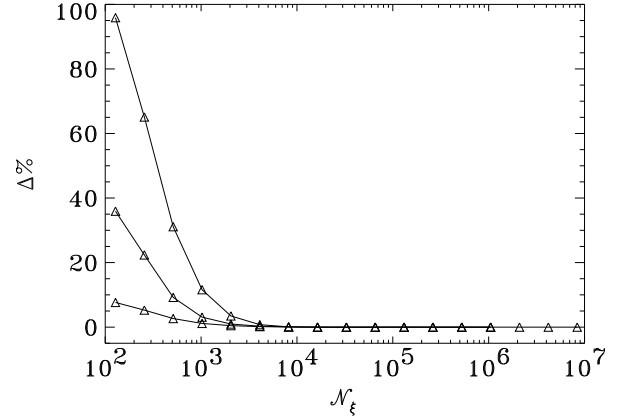}}
\caption{
Shown is the percentual difference between the {\it r.m.s.}\@ of
a ``noise-only'' map obtained with a filter of length $\mathcal{N}_\xi$ and
$\sigma$, the  {\it r.m.s.} obtained for $\mathcal{N}_\xi=\mathcal{N}_d/2$. The
three lines refer, from top to bottom, to $f_k=10$~Hz, $f_k=1$~Hz
and $f_k=0.1$~Hz (the {\sc Planck} goal).
}
\label{rms_vs_minfreq}
\end{figure}

As already discussed in the previous Sections, 
the length of the noise filter is an important issue to
carefully address.
One should expect that in the ``noise-only'' case
the dependence on $\mathcal{N}_\xi$ could be stronger because of the $1/f$ tail in the noise
power spectrum. To address this issue in detail and to 
show the sensitivity of our results to 
$\mathcal{N}_\xi$, the following scheme is employed:
we generate the ``true'', fully correlated 
noise time stream, accordingly to Eq.~(\ref{pdf}).
Then, we do map-making using a noise filter with $\mathcal{N}_\xi 
= \mathcal{N}_d/2$, and
 evaluate the {\it r.m.s.}~of this map,
$\sigma$.\footnote{
Note that we simulate the whole ``noise-only'', $\mathcal{N}_d$ long,
time stream by using Eq.~(\ref{pdf}) and standard FFT techniques. 
This time stream is obviously periodic. Thus, $\mathbf{N}^{-1}$
is strictly circulant even for $\mathcal{N}_d = \mathcal{N}_\xi/2$.
The same would not be true for real data.
}
Then, we generate a set of  maps out of the same 
fully correlated noise time stream, 
but imposing a noise filter length,  $\mathcal{N}_\xi$,
shorter and shorter  w.r.t.\@ 
$\mathcal{N}_d/2$. In Fig.~\ref{rms_vs_minfreq} we 
plot the fractional variation of the map {\it r.m.s.} w.r.t.
$\sigma$  as a function of $\mathcal{N}_\xi$. 
It is clear from the figure that, for the {\sc Planck} Surveyor, we can 
lower $\mathcal{N}_\xi$ by a factor of $\sim$ thousand 
w.r.t.\@  $\mathcal{N}_d/2$ and change the final {\it r.m.s.} 
of the map by much less
than one percent.  This shows that increasing 
$\mathcal{N}_\xi$ above a given threshold does not induce any significant
difference. This result holds even if
we consider knee frequencies significantly
higher than $f_k = 0.1$~Hz: 
in Fig.~\ref{rms_vs_minfreq} we also plot the effect of having
$f_k =  1$~Hz  and  $f_k =  10$~Hz, 
surely two very pessimistic assumptions in the
{\sc Planck} context. 

\begin{figure}
\resizebox{\hsize}{!}{\includegraphics[angle=0]{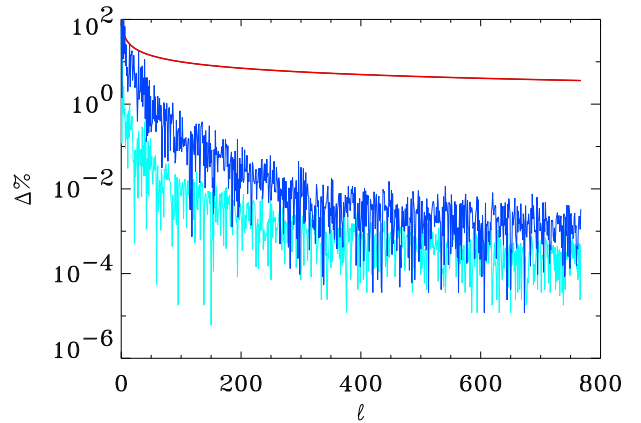}}
\caption{
Shown is the percentual difference between noise-only $C_\ell$'s
obtained for $\mathcal{N}_\xi = 2^{17}$ (light blue curve)
and  $\mathcal{N}_\xi = 2^{13}$ (dark blue curve) w.r.t.\@ 
the power spectrum obtained with the longest possible
filter ($\mathcal{N}_\xi = \mathcal{N}_d/2$).
The red line is the uncertainty due to cosmic variance.}
\label{diff_spe_chunks}
\end{figure}

In principle, the noise properties 
of the {\sc Planck} detectors, as described by Eq.~(\ref{pdf}), 
suggest the existence of spurious correlations in 
the TOD over the whole mission
life-time. In practice, 
the error introduced by neglecting them is negligible, 
at least when the {\it r.m.s.} is used as a figure of
merit. To stress this point even further, we evaluate the angular 
power spectrum from two maps
generated with the same input (\ie\ the same TOD), but using 
noise filters of different lengths.
Again, we plot the percentual difference between the spectrum obtained 
with $\mathcal{N}_\xi =\mathcal{N}_d/2$ and the spectra obtained with 
noise filters shorter by a factor of $64$ and $1024$, respectively. 
Fig.~\ref{diff_spe_chunks} shows that this 
difference is below $\sim 1 \%$
for $\ell \gtrsim 100$,  well under cosmic
variance in the region of interest
for recovering the cosmological parameters.


To conclude this section let us note that
the algorithm discussed here is
intrinsically linear [\cf\ Eq.(\ref{measures})] since 
the map estimator can be written as the ``true'' 
map plus a noise term:
\[ 
{\tilde {\mathbf{m}}} = \mathbf{m} +
(\mathbf{P}^T\mathbf{N}^{-1}\mathbf{P})^{-1}\mathbf{P}^T\mathbf{N}^{-1}
\mathbf{n}.
\]
Our code shares this property.
In fact, if we produce a map from a ``signal-only'' TOD and 
sum it to the map produced from a
``noise-only'' TOD the result is 
virtually indistinguishable (to machine precision) from the map obtained 
by summing the two (``signal-only'' and ``noise-only'') time streams. 

\subsection{Balloon Borne Experiments}\label{Boom_map}

As already mentioned, our code does not take any advantage 
from the specific {\sc Planck} scanning strategy. 
It is then straightforwardly applicable to other one-horned experiments, 
such as BOOMERanG (de Bernardis \etal\  2000).
Here we do not want to enter in any data analysis issue. 
Rather, we want to show the flexibility and
the efficiency of our code to process data coming from an experiment 
completely different from {\sc Planck}.

As far as the signal is concerned, we simulate a theoretical input map
$^9$ pixelized at 1/3 of the BOOMERanG FWHM ($\sim 10^\prime$ @ 150~GHz) 
and properly smeared. 
Knowing its scanning strategy, we extract a typical 
BOOMERanG ``signal-only'' time
stream. As far as the noise is concerned, we generate a
``noise-only'' time stream with the noise properties described in
Eq.(\ref{pdf}) and choosing $A = 150\muKrs$ and $f_k = 0.07$~Hz.
In what follows we consider a noise filter of length $\mathcal N_{\xi} = 2^{17}$.

On the very same line of the previous subsections, we
compare the theoretical input map with $\tilde \mathbf m$, which is  
reconstructed at a
level of accuracy of $10^{-5}$ when the CG solver precision 
is chosen to be $10^{-6}$. The length
of the noise filter does not affect, even in this case, the
final results. 

\begin{figure}
%\resizebox{\hsize}{!}{\includegraphics[angle=+90]{h2645f16.eps}}
%\resizebox{\hsize}{!}{\includegraphics[angle=+90]{h2645f17.eps}}
\caption{Upper panel: map-making output on the simulated BOOMERanG
TOD (signal plus noise). Lower panel: residual noise map obtained
subtracting the input map from the one shown above. The
maps have $\mathcal{N}_p \simeq 6 \times 10^5$.
}
\label{map:boom}
\end{figure}

In the above tests we only consider the 1 degree per second (d.p.s.) 
section of the BOOMERanG scan. However, in order to
facilitate comparisons with {\sc Planck}, 
we truncated the BOOMERanG simulated TOD at the
length of $\mathcal{P}$60. Not surprisingly, our code runs the BOOMERanG
map-making in $\sim 15$ minutes on a 500~MHz Pentium III workstation, while 
the whole  1 d.p.s. BOOMERanG
scan is processed on the same machine in about 20 minutes.

Given these time scales, a parallel machine does not seem necessary 
for the analysis of a single channel. 
However, a parallel environment becomes quite appealing 
if one performs a combined analysis of more than one receiver. 
Also, having a  parallel implementation may be important to perform 
extensive Monte Carlo simulations of the
experiment. Specific applications of this code to BOOMERanG
will be discussed elsewhere.

\section{Summary} 

The purpose of this work was to present a parallel implementation
of a map-making algorithm for CMB experiments. In particular,
we have shown for the first time Maximum Likelihood, 
minimum variance maps obtained
by processing the entire data stream expected from the {\sc Planck} Surveyor,
\ie\ a TOD covering the full mission life span (14 months). Here we restrict
ourselves to the simple case of the {\sc Planck}/LFI 30~GHz channel. However,
the extension of our implemented software to other, higher frequency, channels
is straightforward. At the moment our implementation is limited to
a symmetric antenna beam profile.

The code was shown to scale linearly with the number of time samples
$\mathcal{N}_d$, logarithmically with the noise filter length and
very slowly with the number of pixels $\mathcal{N}_p$, the latter scaling
being mostly influenced by the choice of the preconditioner. 
Furthermore, on a multiprocessor environment the code scales
nearly optimally with the number of processor elements.

The code is extremely accurate and fast. We have shown that it
 recovers the true, input map to a remarkably
high accuracy, over the whole range of considered multipoles. 
Moreover, we run the entire
30~GHz simulated TOD in $\sim$ 90 minutes on an 8 processor job on
a O2K machine, and the BOOMERanG case in only 20 minutes on
a 500 MHz Pentium III workstation. On the latter machine, the {\sc Planck}
case runs in about 15 minutes if we perform the average on circles.

A key problem for a successful map-making pipeline is to obtain a reliable
estimate of the noise properties directly from flight data. It is
straightforward to further iterate the GLS solution implemented here 
to get more accurate estimates 
of the underlying noise. Here we have explicitly
chosen not to do so. Rather, we stop to the {\it zero}\/-th order
solution, ${\tilde \mathbf{n}}$, showing that this is more than enough
for the purposes discussed here.

The possibility of running map-making algorithms on reasonably short
time scales opens up the
possibility of evaluating  the CMB angular
power spectrum via Monte Carlo simulations. We will address
this point in a forthcoming paper, together with the extension to
non-symmetric antenna beam profile.

\begin{acknowledgements} 
We are indebted with C.~Burigana and D.~Maino 
for helping us with the generation of the TOD sets. 
We thank A.~Balbi, P.~Cabella, D.~Marinucci and C.~di~Fiore
for useful suggestions and comments. We also thank
P.~de~Bernardis and the BOOMERanG
collaboration for having provided us with the BOOMERanG scan and 
instrumental performances. PN acknowledges fruitful discussions
with J.~Staren and E.L.~Wright. We acknowledge
use of the HEALPix package and of the FFTW library. 
The supercomputing resources used for this work
have been provided by Cineca.
\end{acknowledgements}

\end{document}